# A LOW POWER HIGH BANDWIDTH FOUR QUADRANT ANALOG MULTIPLIER IN 32 NM CNFET TECHNOLOGY

Ishit Makwana[1] and Vitrag Sheth[2]

[1]Dept. of Electrical & Electronics Engg, Birla Institute of Technology & Science (BITS) Pilani, Pilani, India
ishitmakwana@gmail.com
[2]Hewlett Packard Global Soft India Pvt. Ltd., Bangalore, India
vitragksheth@gmail.com

***ABSTRACT***

*Carbon Nanotube Field Effect Transistor (CNFET) is a promising new technology that overcomes several limitations of traditional silicon integrated circuit technology. In recent years, the potential of CNFET for analog circuit applications has been explored. This paper proposes a novel four quadrant analog multiplier design using CNFETs. The simulation based on 32nm CNFET technology shows that the proposed multiplier has very low harmonic distortion (<0.45%), large input range (±400mV), large bandwidth (~50GHz) and low power consumption (~247µW), while operating at a supply voltage of ±0.9V.*

## 1. INTRODUCTION

As predicted by Moore's law, CMOS manufacturing technology has continued to scale to ever-smaller dimensions now reaching 32nm [1]. At these dimensions several issues arise, such as source to drain tunneling, device mismatch, random dopant fluctuations, mobility degradation, etc. that impact its cost, reliability and performance making further scaling almost impossible. To maintain the trend of Moore's Law and technology's advances in nano-scale regime, novel nanoelectronic solutions are needed to surmount the physical and economic barriers of current technologies.

As one of the promising new transistors, carbon nanotube field-effect transistor (CNFET) overcomes most of the fundamental limitations of traditional silicon MOSFET. The excellent device performance of carbon nanotube (CNT) is attributed to its near-ballistic transport capability under low voltage bias [2]. Intense research on carbon nanotube (CNT) technology has been performed on digital circuit applications such as logic or memory, as well as radio-frequency (RF) devices for analog applications. The potential for high intrinsic device speed [3] and demonstration of CNFETs with a cut-off frequency $f_T$ as high as 80 GHz [4] have indicated that CNT-based devices are well suited as building blocks of future analog and RF circuits.

CNFET structure is similar to a conventional MOSFET except that its semiconducting channel is made up of carbon nanotubes (CNT) as shown in Figure 1. Since the electrons are only confined to the narrow nanotube, the mobility goes up substantially on account of near-ballistic transport as compared to the bulk MOSFET. The near-ballistic transport is due to a limited carrier-phonon interaction because of larger mean free paths of acoustic phonons [5]. Additionally, CNFET shows higher current density, and higher electron mobility of the order of $10^4$–$10^5$ $cm^2/Vs$ [6] compared with $10^3$ $cm^2/Vs$ for bulk silicon. The sizing of a CNFET is equivalent to adjusting the number of tubes. Since the mobility of n-type and the mobility of p-type carriers inside CNTs are identical, the minimum size is 1 for both P-CNFET and N-CNFET [7]. The device is turned on or off by the applied gate voltage. Thus, CNFET is a high quality semiconducting material.

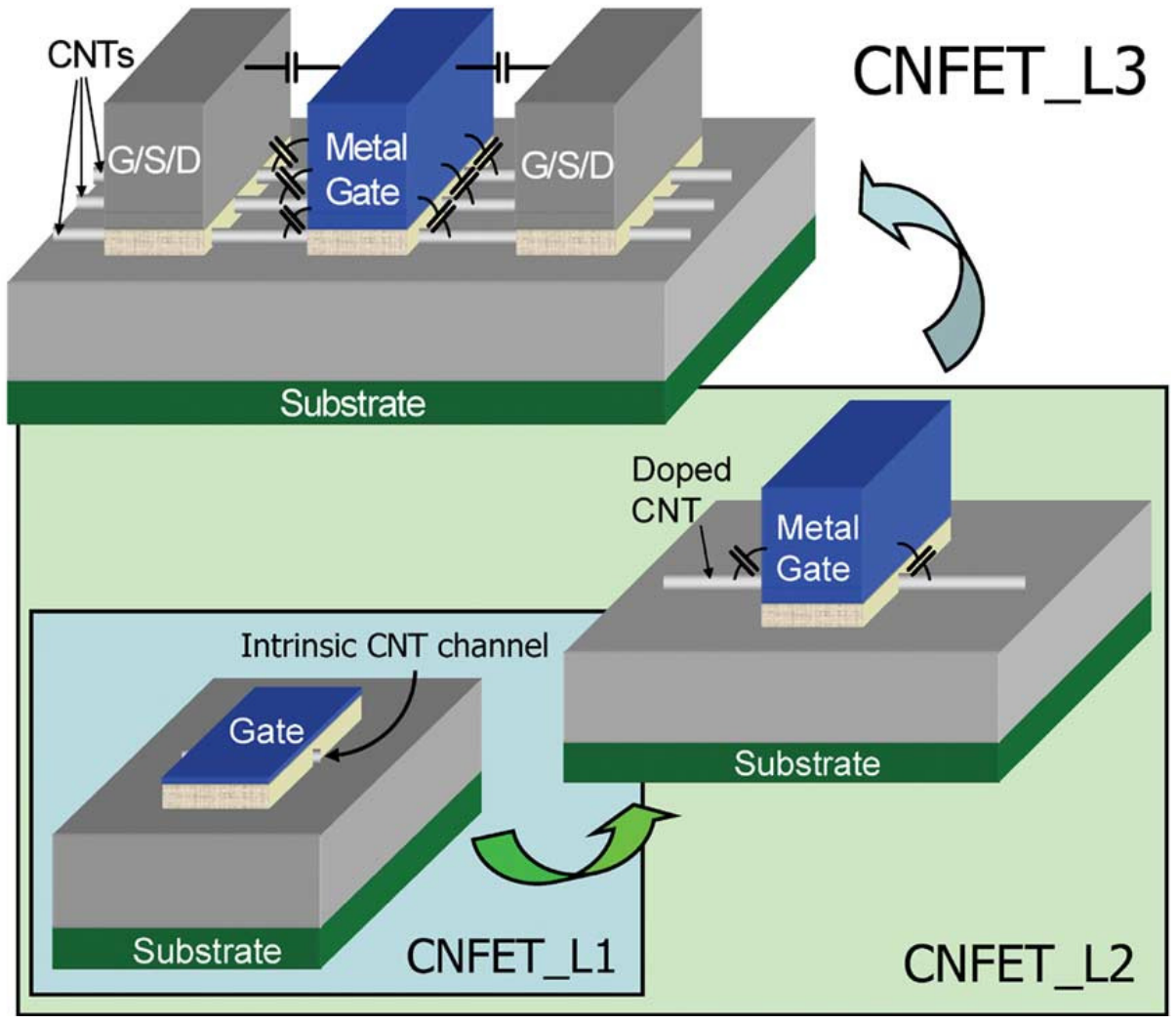

Figure 1. CNFET device structure with multiple nanotubes (CNFET_L3) and with a single tube (CNFET_L2) as illustrated in [2].

The potential of CNFETs for analog circuit design is explored in recent works [7,8]. CNFET also exhibits properties of higher current densities, higher transconductance, lower intrinsic capacitances, as compared to CMOS, which makes CNFET attractive for linear analog circuit applications. It has been demonstrated that CNFETs have the potential to provide higher linearity as compared to silicon or III-V semiconductors [9]. Therefore, low power and high bandwidth analog circuits can be designed based on CNFET.

The four quadrant analog multiplier circuit is a versatile building block used for numerous communication signal processing applications. In contemporary VLSI chips, the analog CMOS multipliers are widely applied in phase-lock loops, automatic variable gain amplifiers, mixers, modulators, demodulators and many other non-linear operations - including division, square rooting, frequency conversion, etc. In most of applications, the desired multiplier's features are good isolation between input-output ports (especially for RF systems), wide input dynamic range, wide bandwidth, symmetric input/output delay, low power dissipation and low voltage supply [10].

Following the early work of Gilbert [11], a variety of multipliers has been designed with different optimization objectives [12-15]. Multipliers have been built around various topologies using Bipolar, CMOS [12,13] and Bi-CMOS [14,15] based circuits. The general idea behind these designs is to utilize electronic devices like BJT or MOS transistors to process the two input signals, followed by a cancellation of errors caused by non-linearity of the devices. MOS transistors are widely used for multiplication process, while differential circuit structure is generally used for nonlinearity cancellation [10].

In this paper, we present a new analog multiplier based on 32nm CNFET technology with emphasis on high linearity, low power consumption, and high bandwidth. We analyze the performance of the multiplier and compare it with recent works.

## 2. ANALOG MULTIPLIER DESIGN

The design of the proposed multiplier is based on the quarter square algebraic identity [16], which is a widely used method for multiplier implementation. In this technique, in the first step the sum and difference of two input signals is performed. Next, the square of both the sum and difference is performed. Finally, the difference between the two square signals is taken as the output. The output signal can be expressed as,

$$V_{out} = [(V_1 + V_2)^2 - (V_1 - V_2)^2] = 4.V_1.V_2 \qquad (1)$$

The proposed multiplier circuit is shown in Figure 3. The multiplier consists of two stages, the first stage is the adder-subtractor circuit and the second stage is the core multiplier circuit, the output of which is taken differentially across two output terminals. The complete multiplier circuit requires the input signal voltages V1 and V2 in both true form and inverted form, i.e. V1, V2 and -V1,-V2.

### 2.1. Capacitive Divider based Adder-Subtractor Circuit

The input stage of the multiplier circuit consists of the adder circuit, which is realized by a capacitive divider based voltage adder/scaling circuit, as shown in Figure 2. A capacitive divider based circuit is chosen for addition of the two input signals, which reduces Total Harmonic Distortion (THD) at the output, reduces power consumption, improves linearity and increases input dynamic range of the multiplier [17]. According to the superposition theorem, the output voltage $V_o$ is the scaled sum of the input voltages $V_1$ and $V_2$. In the complete multiplier circuit shown in Figure 3, the $V_o$ of the adder circuit is connected directly to the gate of one of the NCNFET devices. For example, the $V_y$ voltage at the gate of M3 is equal to $(V_1+V_2)/6$. Similarly, $V_x$ voltage at the gate of M1 is equal to $(V_1-V_2)/6$.

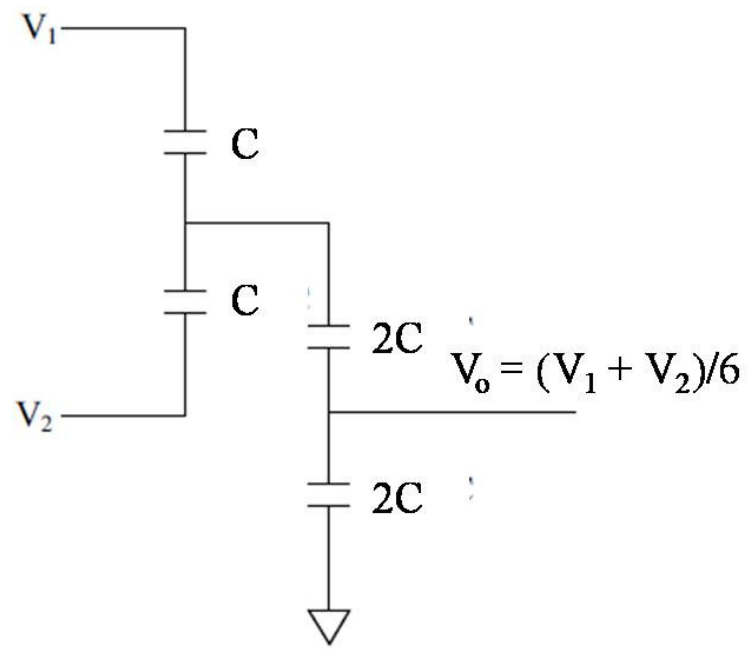

Figure 2. Capacitive adder-subtractor circuit.

At 32nm node, the gate effective capacitance ($C_{eff}$) of CNFET is found to be approximately 200aF/FET [18]. Therefore, to minimize the loading effect of input stage on the gate of CNFET, the value of C in capacitive adder is chosen to be 5fF, which is very large compared to $C_{eff}$ of CNFET. Here the matching of capacitance values is essential for obtaining distortion-free output. The capacitive adders/dividers are preferred to resistive adder/dividers, as getting precise capacitive ratios is easier in CMOS manufacturing process as compared to resistance ratios.

Hence, the use of capacitive adder/divider circuit reduces mismatch effects and reduces harmonic distortion.

## 2.2. The Multiplier Core

In this paper, the high performance Stanford standard model of CNFET have been used for simulation and analyses of analog multiplier circuit [19] which accounts for the practical non-idealities in CNFET. The model also predicts the dynamic and transient performance of CNFET with more than 90% accuracy [20]. The initial design parameters of CNFET topology was set to their most practical values and then optimized for the multiplier circuit.

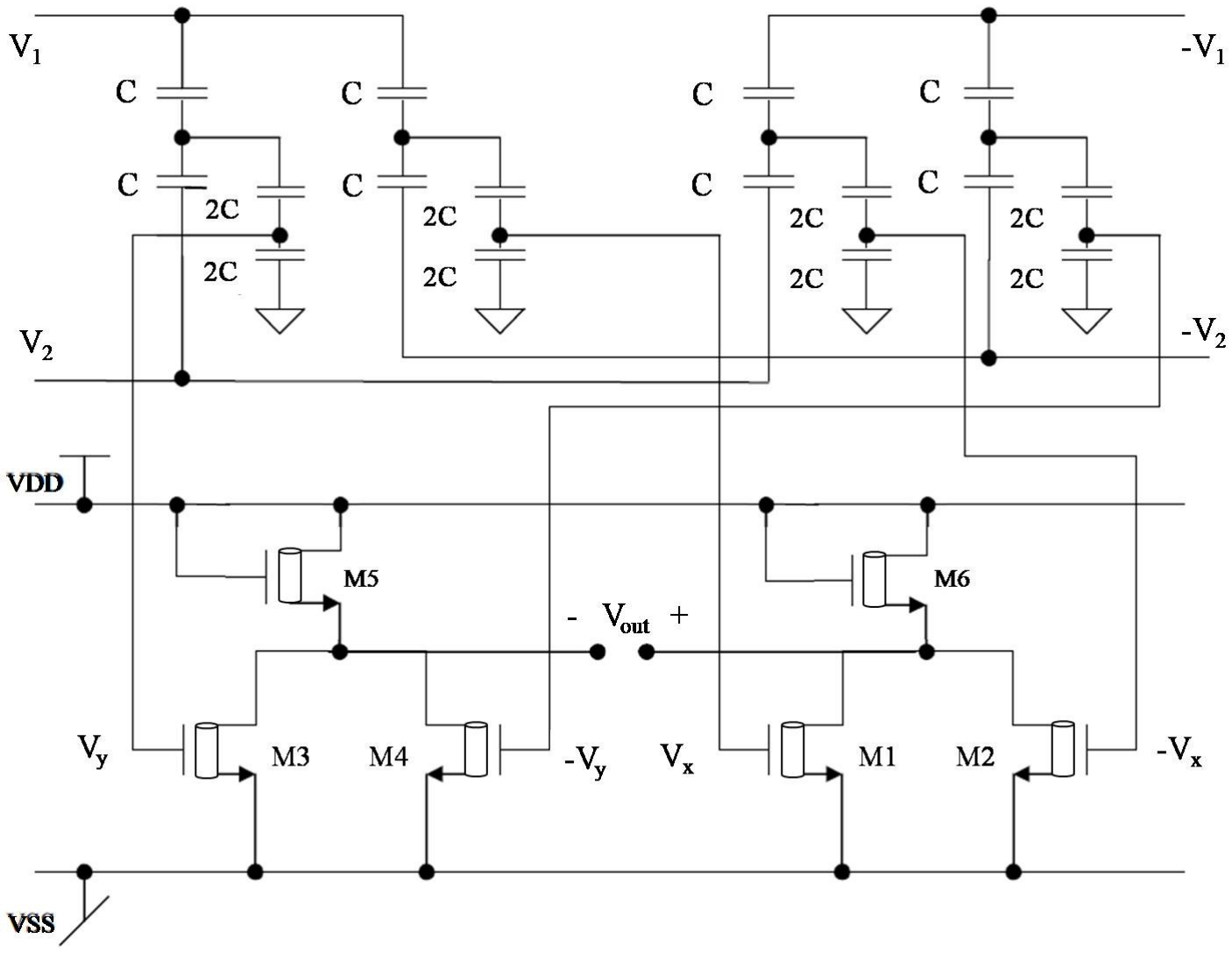

Figure 3. Proposed multiplier circuit.

The relatively larger value of $g_m$ in CNFET as compared to CMOS, makes it useful for analog applications such as amplification and modulation.

$$g_m = C_g' \frac{abs(V_g - V_{g0})\mu}{L} \qquad (2)$$

The limiting value of $g_m$ for a ballistic nanotube transistor has been shown to be 150µA/V [2]. It is also found that there is a direct relationship between CNFET diameter, band-gap energy $E_g$ and threshold voltage $V_{th}$ of intrinsic CNT channel. It is given by the following equations [2][20]:

$$E_g = \frac{0.84eV}{D_{CNT}} \qquad (3)$$

$$V_{th} = \frac{E_g}{2e} = \frac{0.577aV_{\pi}}{D_{CNT}} \qquad (4)$$

$$D_{CNT} = a\frac{\sqrt{n^2 + nm + m^2}}{\pi} \qquad (5)$$

The parameters, 'a' (~2.49 Å) is the carbon to carbon atomic distance, $V_\pi$ (~3.033 eV) is the carbon p-p bond energy in the tight bonding model, 'e' is the unit electron charge and (n,m) is the chirality. The diameter of tube $D_{CNT}$ is the main parameter that directly affects the transconductance in a CNFET. The barrier height at the S/D contact ($E_{f0}$), chirality (n,m) and oxide thickness affect the gain and noise performance of CNFET. The power consumption of the amplifier increases due to smaller ($E_{f0}$) as CNT become more conducting, increasing the on-current. Moreover, the bandwidth also improves with increase in diameter of CNT, which is due to enhanced screening between adjacent channels [7].

The inputs to multiplier are $V_x=(V_1-V_2)/6$ and $V_y=(V_1+V_2)/6$. Here, the scaling factor of 1/6 ensures that $V_{gs}$ at gate of CNFETs is always greater than the threshold voltage $V_{th}$ of NCNFET when input voltages are in the range ±400mV. The transistors M5 and M6 have twice the number of CNT as compared to M1-M4 transistors since the current through M5 and M6 is twice that of M1-M4. Moreover, the transistors M5 and M6 are biased in saturation region such that the DC node voltage at output terminals $V_{o1}$ and $V_{o2}$ is 0V. Thus all transistors in the multiplier operate in saturation region. The non-linearity effect in the output is cancelled by matching pairs M3-M4 and M1-M2 and taking differential output across the drain terminal of the two pairs. Thus, the above configuration of the circuit helps achieve high linearity multiplication across wide input voltage range.

## 3. Simulation Results

The proposed multiplier circuit has been simulated with Synopsys HSPICE simulator using the 32nm CNFET SPICE model from Stanford University at a supply voltage of ±0.9V [19]. The following CNFET technology parameters are used for each CNFET in the simulation of multiplier:

Table 1. Technology Parameters of CNFET used in the proposed circuit

| Parameter | Value |
|---|---|
| Physical length of channel | 32.0 nm |
| Length of doped CNT source/drain extensions | 32.0 nm |
| Fermi level of doped source/drain CNT regions | 0.6 eV |
| Work function of source/drain metal contact | 4.6 eV |
| CNT Work function | 4.5 eV |
| Mean free path in Intrinsic CNT | 200.0 nm |
| Mean free path in p+/n+ doped CNT | 15.0 nm |
| Chirality of the tube | (25,0) |
| Sub-lithographic Pitch | 6.4 nm |
| Pitch (distance between two nanotubes) | 20.0 nm |
| Dielectric constant of high-K gate oxide | 16.0 |
| Thickness of high-K gate oxide | 4.0nm |

The sizing of transistors is performed according to the magnitude of DC currents as mentioned above. The number of tubes for *M1-M4* is 1 and number of tubes for *M5-M6* is 2. The major parameter which affects the performance of multiplier is diameter of CNT ($D_{CNT}$). It is observed that with increase in $D_{CNT}$ the bandwidth, linearity range and power consumption of the multiplier increases. The optimum diameter is found to be 1.98 nm, corresponding to a chirality of (25,0), for achieving high linearity and wide input range, while the power consumption of the multiplier is kept low by keeping minimum number of tubes per device.

The performance of the multiplier is evaluated on the parameters of linearity, input range, power consumption, frequency range and noise. Depending upon applications, some of them can be more important than others. There is a trade-off between certain parameters. For instance, as the power consumption of multiplier is decreased, the linearity range and bandwidth also decreases. Hence, a reasonable trade-off is required to achieve low power consumption and good linearity.

### 3.1. DC Transfer Characteristics

The DC transfer characteristics of the circuit with $V_1$ and $V_2$ as the two inputs are shown in Figure 4. The $V_{out} = V_{o2}-V_{o1}$ swings between -4.5mV to +4.5mV for the input range of $\pm$0.4V. In the simulation, 'in1' is the variable for input $V_1$ depicted on X-axis, and $V_2$ is varied from -0.4V to 0.4V with 50mV step size. The circuit is symmetric with respect to both input, therefore the DC characteristics remain same. At the operating point, the DC node voltage at $V_{o1}$ and $V_{o2}$ is nearly 0V (~0.677mV). The quiescent power consumption of the overall circuit is only 246.9μW.

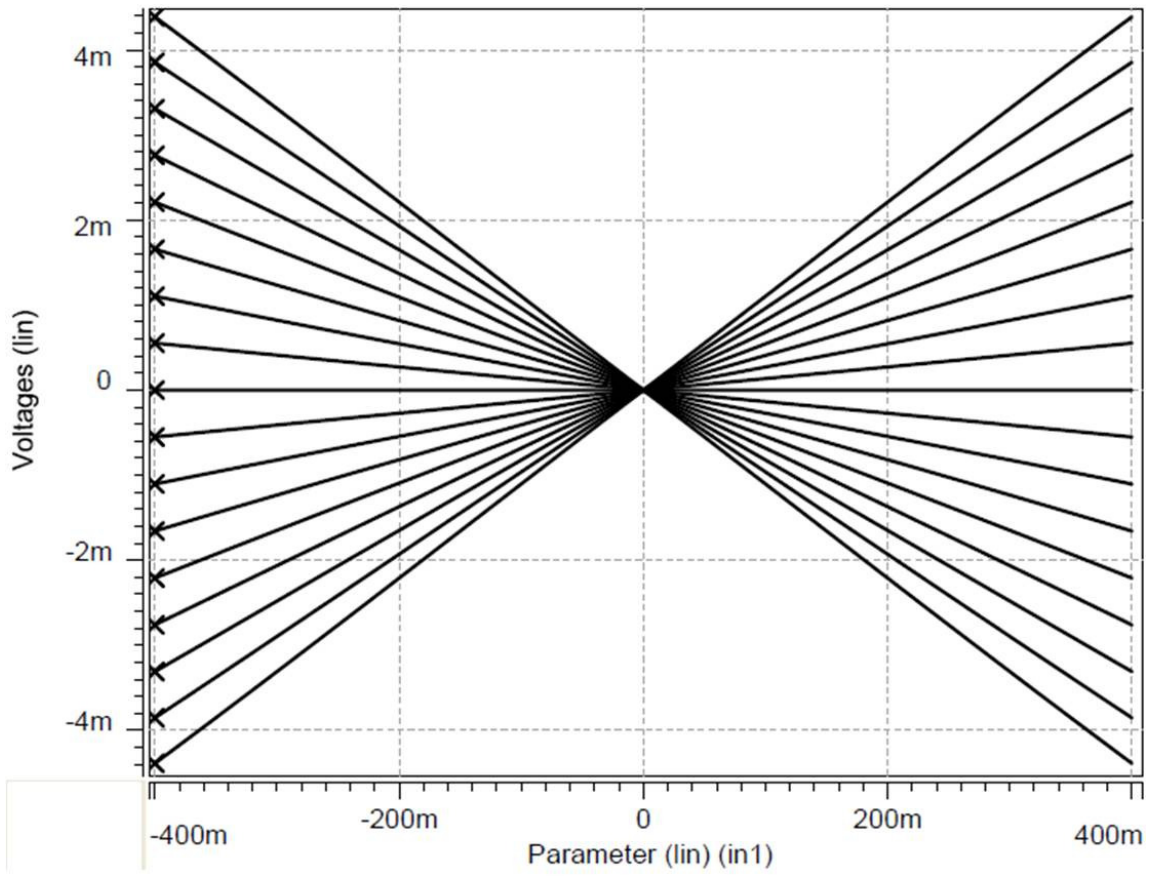

Figure 4. DC Transfer Characteristics of multiplier

### 3.2. Harmonic Distortion Analysis

The harmonic distortion analysis is performed in frequency doubler configuration of the multiplier. In this configuration, a 0.4V amplitude sinusoidal signal of 1MHz frequency is given as input to both $V_1$ and $V_2$. Figure 5 depicts the transient response of the multiplier. The output signal is sinusoid with fundamental frequency 2MHz. Figure 6 shows the simulated output spectrum of the output signal obtained by Fourier analysis of the transient response in frequency doubler configuration. The output component at 2MHz is having the highest magnitude as expected. It is observed that the harmonic frequency components are at least 3 orders of magnitude less than the fundamental. Figure 7 shows the variation of Total Harmonic Distortion (THD) in percentage of the output waveform with different $V_1$ and $V_2$ values. Here both $V_1$ and $V_2$ have same input frequency of 1MHz. It is found that the multiplier exhibits very low THD (<0.45%) for a wide input range of ±400mV.

### 3.2. Frequency Response

Figure 8 shows the frequency response of the proposed multiplier in frequency doubler configuration. The simulation is done with a sine wave of 0.2V peak value given as input to both $V_1$ and $V_2$. The -3dB bandwidth is approximately 49.88GHz. The large bandwidth of the multiplier is mainly due to the fact that the intrinsic capacitance of CNFET is much less as compared to MOSFET [18]. Moreover, the gain of the circuit is much lesser than unity, while the structure of the multiplier is similar to a common source amplifier. The bandwidth of the proposed multiplier is close to the cut-off frequency $f_T$ (~80GHz) of the single-walled carbon nanotube common source amplifier as reported in [9].

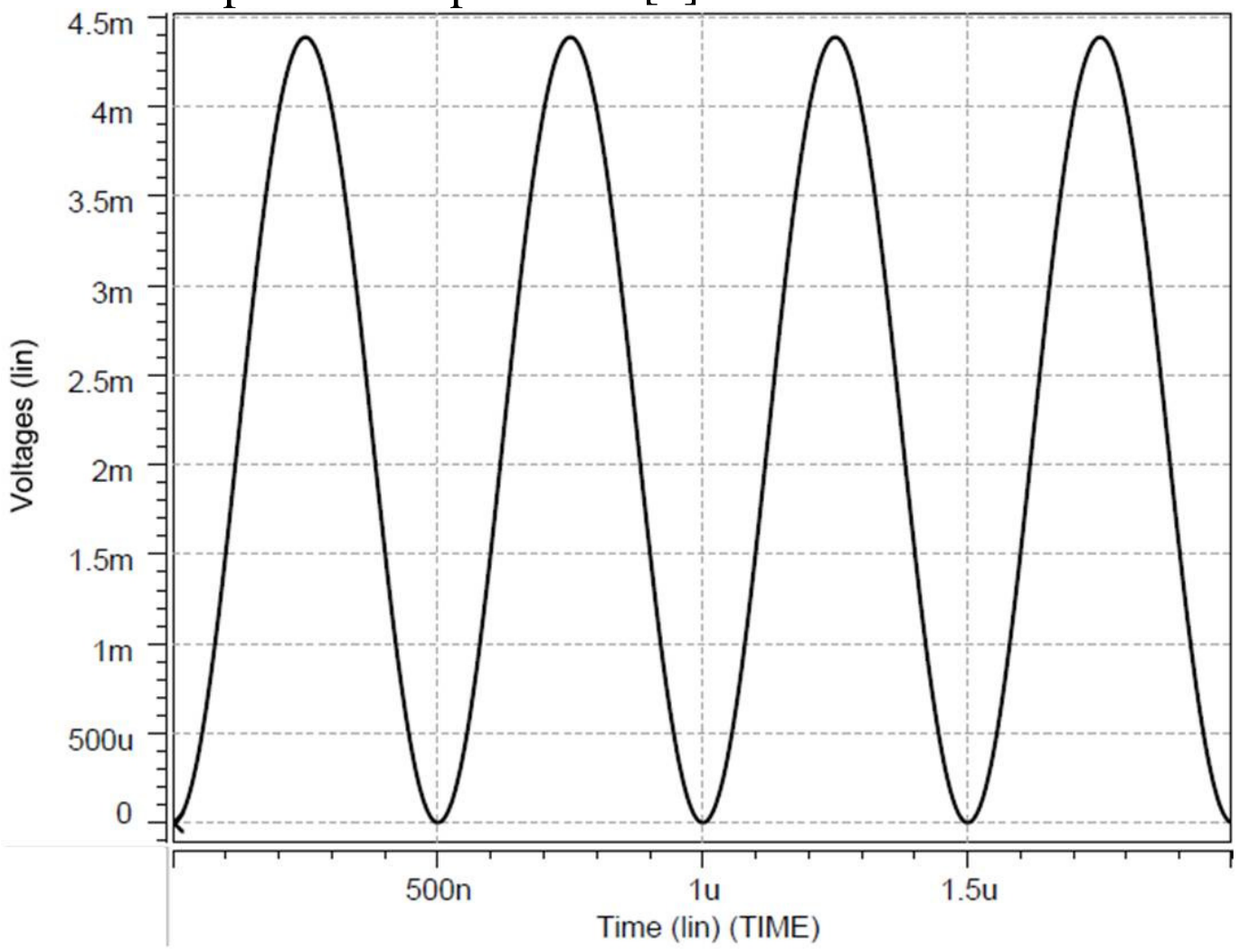

Figure 5. Transient response of the multiplier in frequency doubler configuration with 400mV 1MHz AC input.

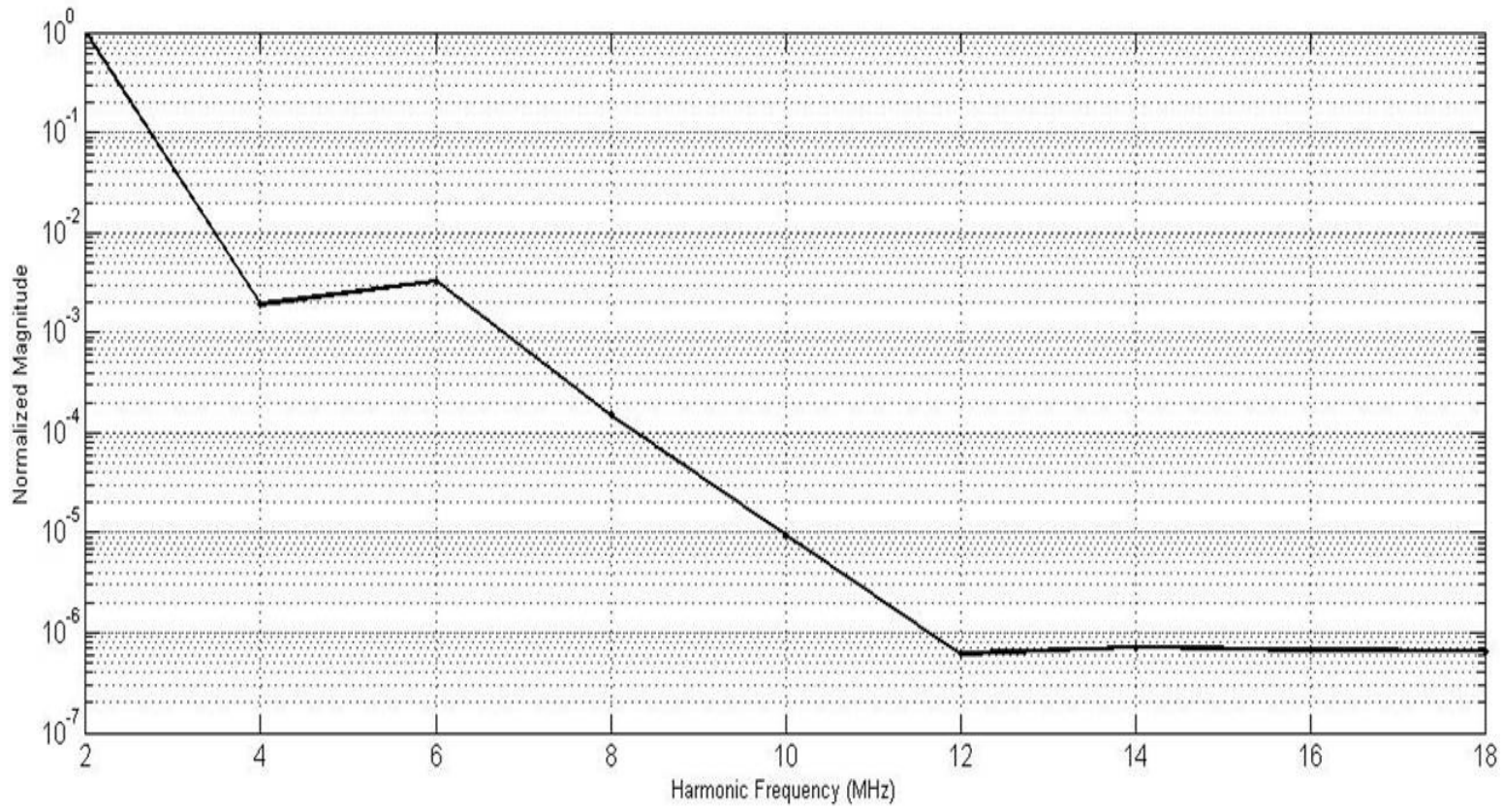

Figure 6. Simulated output spectrum of the multiplier in frequency doubler configuration for 0.4V 1MHz AC input signal.

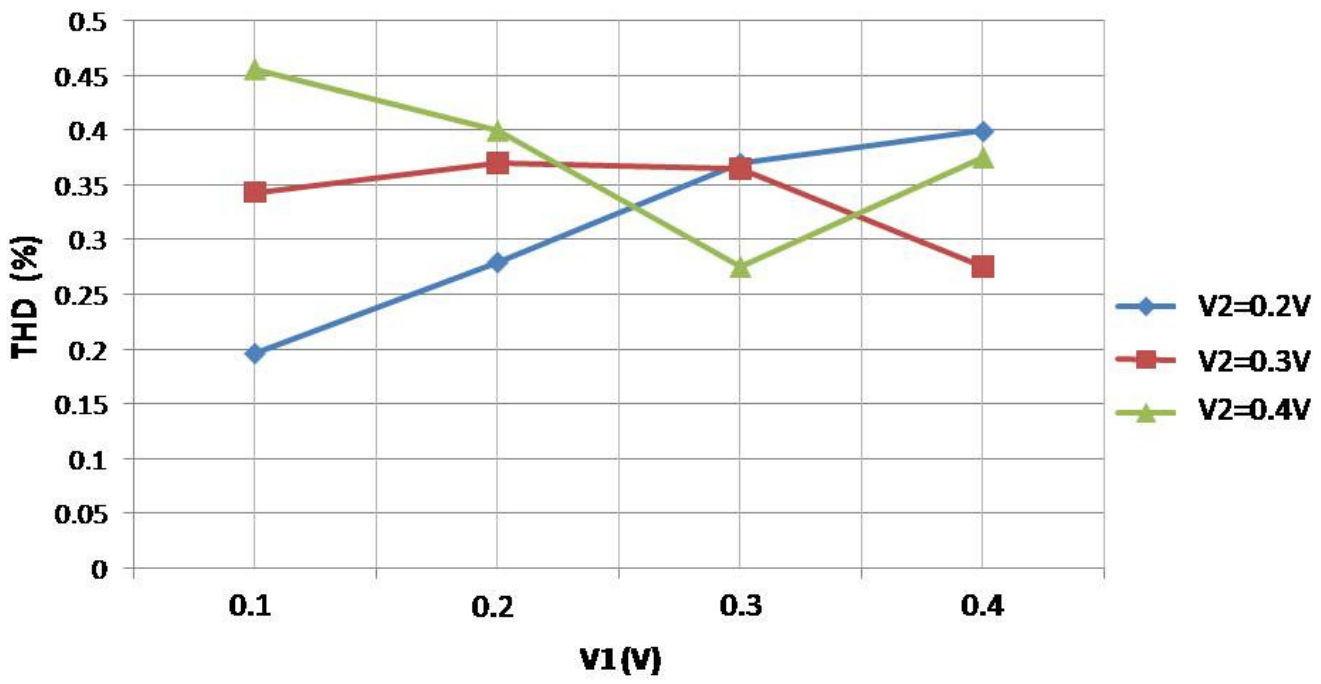

Figure 7. Variation of output THD (%) with different $V_1$ and $V_2$ input voltages.

.

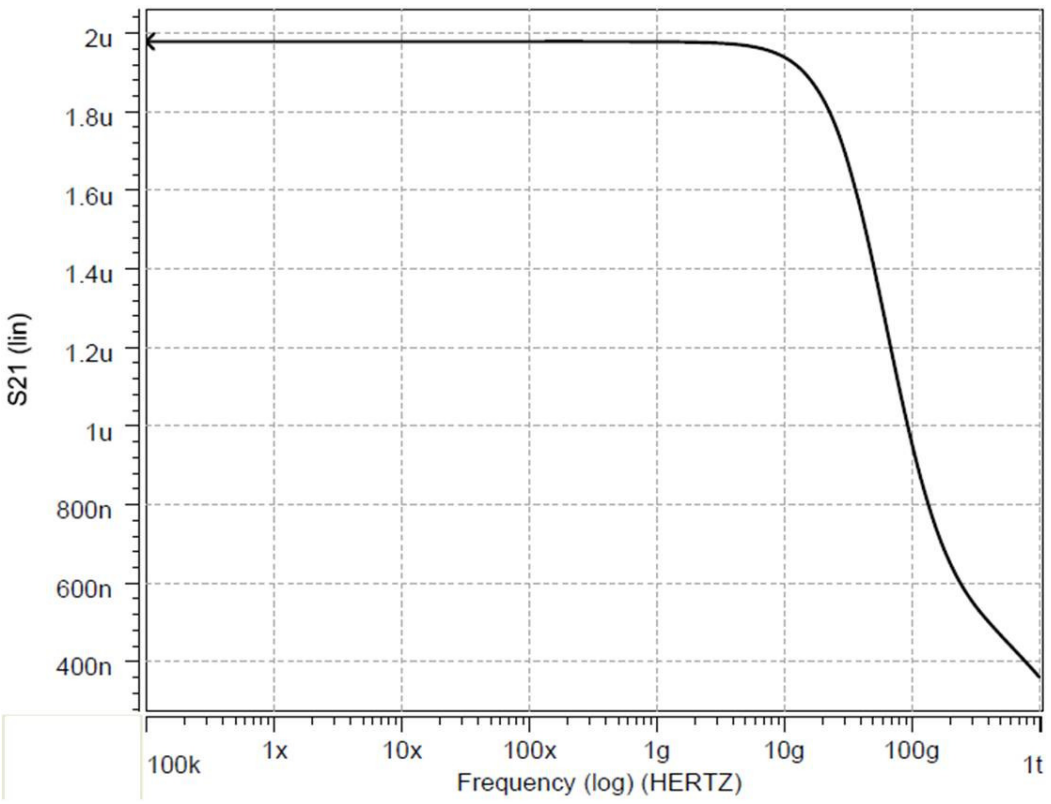

Figure 8. Frequency Response: output magnitude plot for 0.2V AC signal input.

### 3.4. Noise Analysis

The equivalent input noise and the equivalent output noise are plotted in Figure 9 and Figure 10 with input frequency varying from 100 KHz to 1 THz. The simulation is done with the input sine wave value of 0.4V peak in the frequency doubler configuration of the multiplier. The Y-axis is in Volts/√Hz. It is apparent that significant noise suppression is achieved, which can be attributed to the less count of devices used in the circuit.

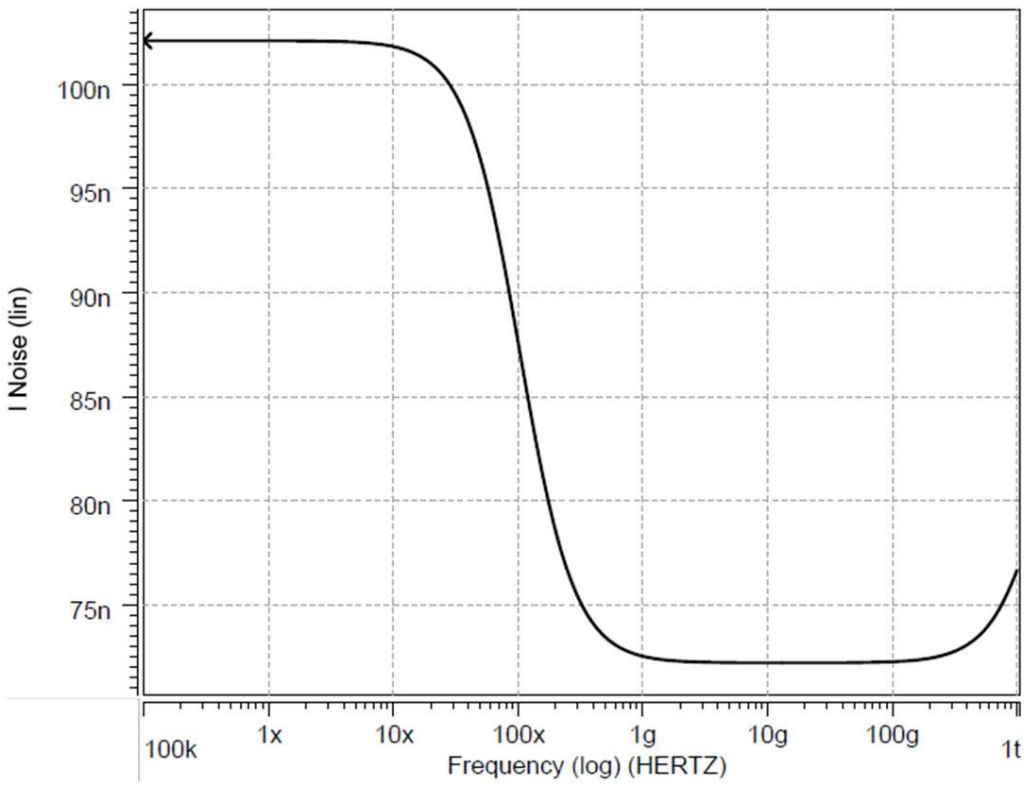

Figure 9. Equivalent Input Noise Plot (in V/√Hz) for input 0.4V AC signal.

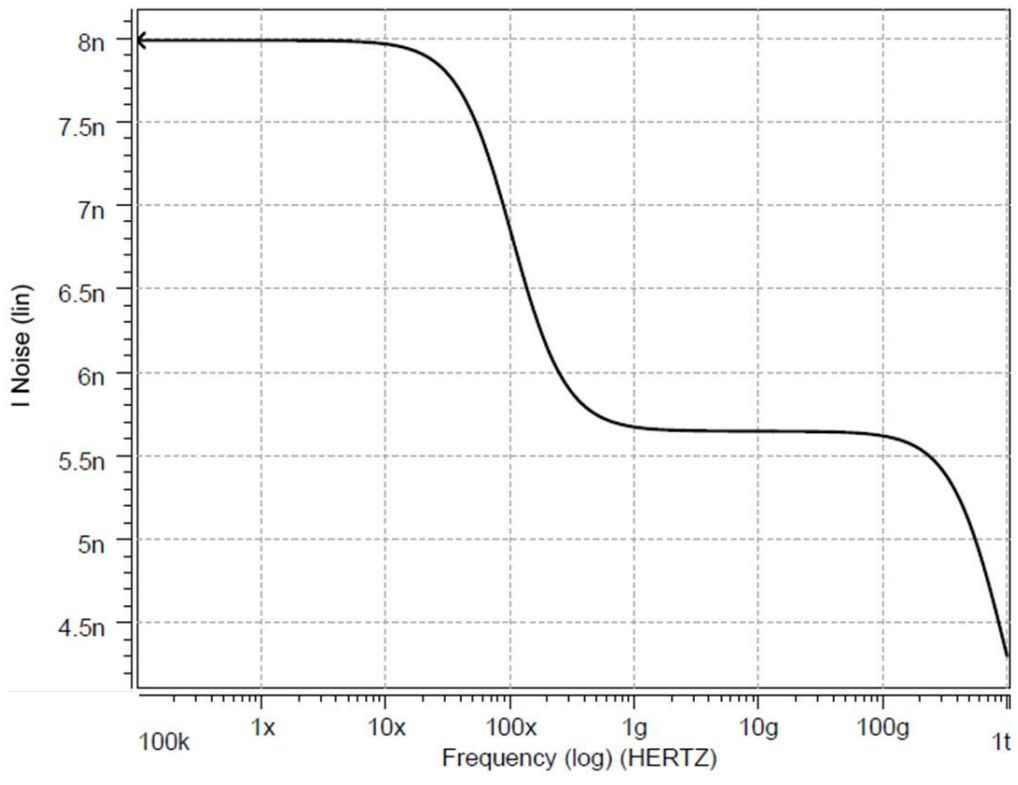

Figure 10. Equivalent Output Noise Plot (in V/√Hz) for input 0.4V AC signal.

### 3.5. Amplitude Modulation

An application of analog multiplier is amplitude modulation. Figure 11 shows the modulation performance of proposed circuit with sinusoidal inputs, $V_1$ is 200mV at 10MHz and $V_2$ is 50mV at 100 MHz. The modulation performance is also simulated at a high frequency of 10 GHz as shown in Figure 12.

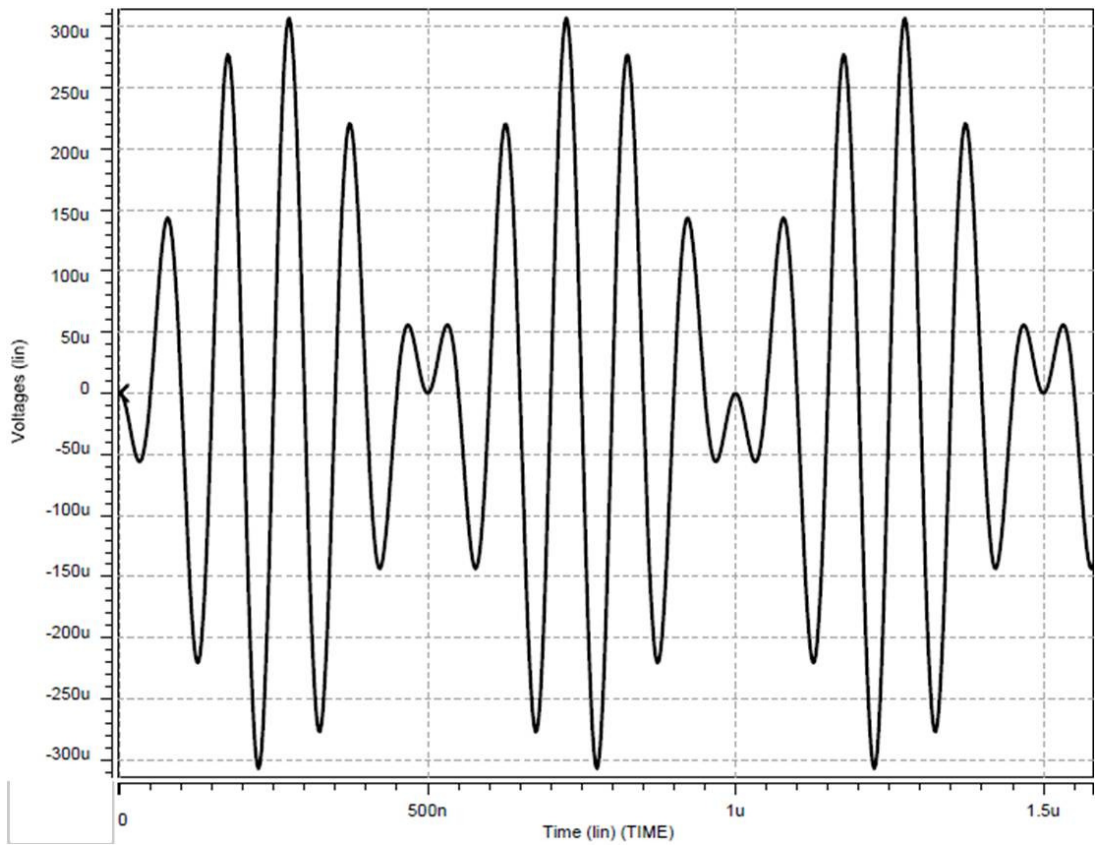

Figure 11. Amplitude Modulation output for inputs 200mV 10MHz carrier and 50mV 1MHz modulating signal.

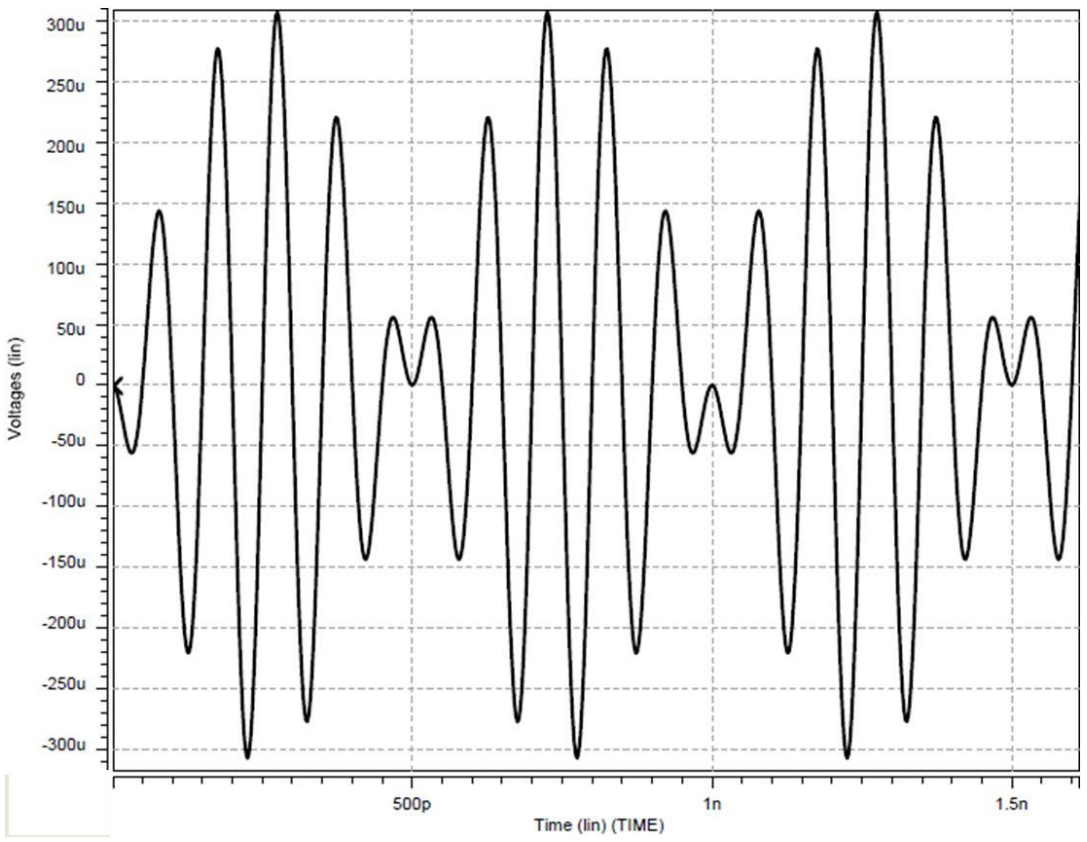

Figure 12. Amplitude Modulation output for inputs 200mV 10GHz carrier and 50mV 1GHz modulating signal.

### 3.6. Comparison with earlier works

The Table 2 summarizes a comparison of the proposed CNFET based multiplier circuit with similar prior work on four quadrant multipliers in CMOS technology. It is observed that the proposed multiplier is better than other CMOS based structures in terms of power consumption, input range, linearity, bandwidth and transistor count.

## 4. CONCLUSIONS

A novel fully differential four quadrant analog multiplier using carbon nanotube FETs is proposed. The proposed multiplier is a two stage structure based on quarter square technique of multiplication. The circuit is designed for low power and high bandwidth operation. A detailed analysis of the circuit behaviour has been performed. High linearity and good noise performance has been achieved by well designed structure of multiplier and fine-tuned parameters of CNFET. The performance of the proposed multiplier is also compared with similar works on four-quadrant multipliers in CMOS technology.

Table 2. Comparison of the proposed multiplier with previous works

| Factor | This work | Ref [21] | Ref [22] | Ref [23] | Ref [24] | Ref [25] |
|---|---|---|---|---|---|---|
| Technology Node | CNFET 32nm | CMOS 0.18μm | CMOS 0.5μm | CMOS 0.8μm | CMOS 0.35μm | CMOS 0.35μm |
| Supply Voltage | ±0.9V | ±1V | ±2.5V | +1.2V | +1.8V | +1.5V |
| Input Range | ±400mV | ±100mV | ±1V | ±250mV | ±400mV | ±400mV |
| THD % at 1MHz | ≤0.45 | ≤1.0 | ≤0.85 | ≤1.1 | ≤1.0 at 25KHz | ≤1.0 |
| Power | 246.9 μW | 588μW | 3.6mW | 2.76mW | 200μW | 46.4 μW |
| -3dB Bandwidth | 49.88GHz | 3.96GHz | 120MHz | 2.2MHz | 10MHz | 95MHz |
| Transistor Count | 6 | 36 | 27 | 12 | 10 | 10 |

## Authors

Ishit Makwana received the B.Tech. Degree in Electronics & Communication from Institute of Technology, Nirma University, Ahmedabad. He is currently working toward the Master of Engineering Degree in Microelectronics, at Birla Institute of Technology & Science (BITS) Pilani, Rajasthan. His research interests include design and analysis of nanoscale devices and circuits.

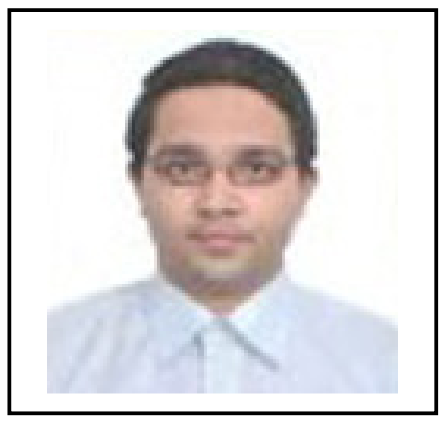

Vitrag Sheth received the B.Tech. Degree in Electronics & Communication from Institute of Technology, Nirma University, Ahmedabad. He is currently with Hewlett Packard Global Soft Pvt. Ltd., Bangalore. His research interests include analog and digital VLSI design.

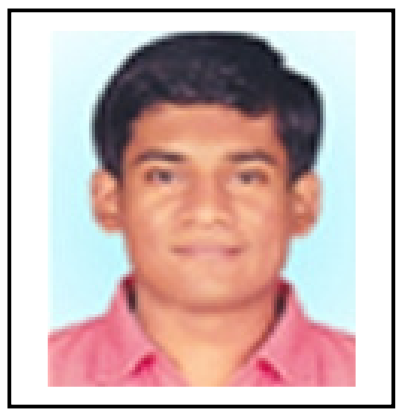